%% file: main.tex
\documentclass[letterpaper, 10 pt, conference]{ieeeconf}  % Comment this line out
\IEEEoverridecommandlockouts                              % This command is only
\usepackage{graphicx}
\usepackage[spanish]{babel}
\usepackage[utf8]{inputenc}
\usepackage{hyperref}
\usepackage{multirow}
\usepackage{slashbox}

\title{\LARGE \bf
Performance Evaluation of Container-based Virtualization for High Performance Computing Environments
}

\author{Carlos Arango$^{1}$, Rémy Dernat$^{3}$, John Sanabria$^{2}$% <-this % stops a space
\thanks{$^{1}$ Facultad de Ingeniería, Escuela de Ingeniería de Sistemas y Computación,
        Universidad del Valle, Colombia
        {\tt\small carlos.arango.gutierrez@correounivalle.edu.co}}%
\thanks{$^{2}$P. Facultad de Ingeniería, Escuela de Ingeniería de Sistemas y Computación,
        Universidad del Valle, Colombia
        {\tt\small john.sanabria@correounivalle.edu.co}}%
\thanks{$^{3}$ 
ISEM, CNRS, Univ. Montpellier, IRD, EPHE, Montpellier France
        {\tt\small remy.dernat@umontpellier.fr}}%
}

\begin{document}
\maketitle
\thispagestyle{empty}
\pagestyle{empty}

\input{abstract}

\input{introduction}

\input{costechnologies}
\input{methodology}

\input{related_work}
\input{conclusions}
\input{acknowledgments}

\addtolength{\textheight}{-12cm}   % This command serves to balance the column lengths

%%%%%%%%%%%%%%%%%%%%%%%%%%%%%%%%%%%%%%%%%%%%%%%%%%%%%%%%%%%%%%%%%%%%%%%%%%%%%%%%

%%%%%%%%%%%%%%%%%%%%%%%%%%%%%%%%%%%%%%%%%%%%%%%%%%%%%%%%%%%%%%%%%%%%%%%%%%%%%%%%

%%%%%%%%%%%%%%%%%%%%%%%%%%%%%%%%%%%%%%%%%%%%%%%%%%%%%%%%%%%%%%%%%%%%%%%%%%%%%%%%

%%%%%%%%%%%%%%%%%%%%%%%%%%%%%%%%%%%%%%%%%%%%%%%%%%%%%%%%%%%%%%%%%%%%%%%%%%%%%%%%

% main.bib 
\bibliographystyle{plain}
\bibliography{main.bib}

\end{document}

%% file: abstract.tex
\begin{abstract}
Virtualization technologies have evolved  along with the development of computational environments since virtualization offered needed features at that time such as isolation, accountability, resource allocation, resource fair sharing  and so on. 
Novel processor technologies bring to commodity computers the possibility to emulate diverse environments where a wide range of computational scenarios can be run. 
Along with processors evolution, system developers have created different virtualization mechanisms where each new development enhanced the performance of previous virtualized environments. 
Recently, operating system-based virtualization technologies  captured the attention of communities abroad (from industry to academy and research) because their important improvements on performance area.

In this paper, the features of three container-based operating systems virtualization tools (LXC, Docker and Singularity) are presented.  
LXC, Docker, Singularity and bare metal  are put under test through a customized single node HPL-Benchmark and a MPI-based application for the multi node testbed. Also the disk I/O performance, Memory (RAM) performance, Network bandwidth and GPU performance are tested for the COS technologies vs bare metal.  
Preliminary results and conclusions around them are presented and discussed.
\end{abstract}


%% file: introduction.tex
\section{Introduction}

Computational tools are key elements in the development of differents areas of knowledge such as industry, research and academy. 
Simulations and modeling are important computational techniques used to reduce waiting times and money budgets bringing novel and effective solutions to challenging problems. 

New solutions usually required to be obtained through processor-intensive applications which demand specialized infrastructures to perform on acceptable time.
High Performance Computing (HPC) is the name given to those processor-intensive applications to take advantage of massive parallel infrastructures known as computational clusters.

Computational clusters fulfill most of the processor-intensive applications requirements, tackling novel problems and presenting foreseeable solutions.
However, more challenging problems surpass the capacity of one computational cluster and federations of scattered clusters are necessary to meet the needs of these problems.
Those federations of clusters are known as Grid systems.

Grid systems offer virtual organizations which integrate users and computational resources abroad.
Thus, multiple virtual organizations are consolidated world wide tackling diverse problems (e.g. cancer cure, search for fundamental particles and sequencing genomes, among others) then requiring diverse services and applications.

This babel of tools presents a challenging problem for system administrators who have to deal with library versions, dependencies and software compatibility.

Virtualization is not a new technology \cite{rosenblum2005virtual} but it has been recently reactivated because of the advantages that it exhibits.
Nowadays, off the shelf processors incorporate optimized virtualization instructions to support the deployment of secure and isolated computational environments bringing power efficient computational environments able to run several services in one single box\cite{uhlig2005intel,xavier2013performance}.

Cloud computing then emerges as a new infrastructure to borrow the best of Grid Computing and Virtualization in such a way that several users and projects are able to share computational resources in an isolated fashion,\cite{buyya2008market}.
Cloud computing additionally exhibits other characteristics such as ubiquitous access, scalability on-demand and pay for consumed resources, \cite{Mell:2011:SND:2206223}.
Infrastructure, development platforms and software services have took advantage of it and a new economy around to Cloud computing infrastructures have emerged \cite{foster2008cloud}.
%These heterogeneity of requirements rely on virtualized resources which run in a isolated fashion avoiding compatibility problems found in Grid environments\cite{buyya2008market}.
%Along with virtualization instructions incorporated into commodity hardware 
%Recently, virtualization technologies have exhibited an important development thanks to important advances on processor architectures which optimize the management and execution of processor instructions for virtualized contexts \cite{uhlig2005intel}. 
%The main benefits of virtualization include hardware independence, availability, isolation and security.\cite{xavier2013performance}. 
%Hardware advances then leverage the consolidation of virtualization tools to take advantage of new functionalities and bring novel computational scenarios where new applications are impacting our daily life, e.g. social networks and cloud computing\cite{buyya2008market}.

However, HPC is one of the few scenarios where Cloud computing has fall short on providing the performance expected by HPC applications. 
%See also Amazon HPC and CnfCluster
Although important milestones have been reached in the virtualization context and some cloud providers make available tailored virtual computational tools, the performance of virtualized contexts are very slow when they are compared with their bare metal counterpart \cite{Jackson:2010:PAH:1931470.1931896}.

Many scientific and academic applications taking advantage of native and optimized processor instructions which are penalized when they are executed on top of hypervisor tools.
Hypervisors present a simplified view of the native hardware to the virtual machines then they can barely access to the optimized set of instructions of actual processors.

\begin{figure}[t]
\begin{center}
\includegraphics[width=0.4\textwidth]{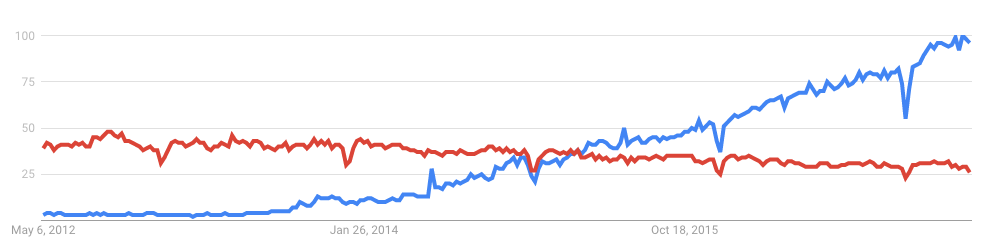}
\caption{\label{fig:intovtime} Container (blue) vs Virtual machines (red) interest over time. \cite{gtrends}}
\end{center}
\end{figure}

%Virtualization has been gaining significance on cloud computing in the past years \cite{felter2015updated}. But using virtual machines (VMs) for parallel computing isn’t an attractive option for many users, as the performance loss is too important. Though, a VM gives users freedom of choosing an operating system, software stack and for the scientific community, they supply repetitive and reproducible ability for their work. With VMs, the researchers can ensure that they work will run on every single machine reviewers will test it. And thus avoiding the common issue of ``it worked on my workstation'' but not on the reviewer Cluster or Workstation.

%%%%%%%%%%%%%%%%%%%%%%%%%%%%%%%%%%%%%%%%%%%%%%%%%%%%%%%%%%%%%%%%%%%%%%%%%%%%%%%%
%The well-known solutions for cloud computing, both commercial (Amazon Cloud\footnote{https://aws.amazon.com/}, Google Cloud\footnote{https://cloud.google.com/}, Microsoft Azure\footnote{https://azure.microsoft.com}, OpenShift\footnote{https://www.openshift.com/}, etc.) and open-source (OpenStack\footnote{https://www.openstack.org/}, CloudStack\footnote{https://cloudstack.apache.org/}...) provide platforms for running a single VM or multiple VMs. This approach comes with drawbacks, which include high performance overhead when running inside a VM \cite{xavier2013performance,ostermann2009performance}, increased complexity of cluster management, as well as the need to learn new tools and protocols to manage the clusters.

An alternative approach to the hypervisor-based solution to virtualized environments has gained traction and attention. 
Containers\cite{linuxcontainers} subtract the hypervisor layer of the virtualization equations and relies on namespaces and cgroups in order to provide isolation and accounting of the consumed resources by the container instances.
%a new player has come to the game of distributed systems, adopting the name of containers \cite{linuxcontainers}. Linux Containers make use of Linux Namespaces to manage resource isolation for single processes, and the Linux cgroups to manage resources for a group of processes. Containers technologies allows the isolation and usage of system resources, such as CPU and memory, for a group of processes.

Then, the rapid development of container-based technologies is getting attention of Internet users because containers accelerates the development process, eases distribution and deployment of applications, Figure \ref{fig:intovtime}. 
Leaders of such development are Docker\footnote{http://www.docker.com} \cite{merkel2014docker} and Linux Containers (LXC\cite{helsley2009lxc}). 
Nevertheless its implications for scientific computing including HPC are still on doubt. 
%fig \ref{fig:intovtime} shows the scientific interest on Docker vs Virtual machines over time from \cite{gtrends}. Numbers represent search interest relative to the highest point on the chart for the given region and time. A value of 100 is the peak popularity for the term. A value of 50 means that the term is half as popular. Likewise a score of 0 means the term was less than 1$\%$ as popular as the peak.
%%%%%%%%%%%%%%%%%%%%%%%%%%%%%%%%%%%%%%%%%%%%%%%%%%%%%%%%%%%%%%%%%%%%%%%%%%%%%%%%

Containers are proving to be an extremely valuable technology for science delivering portability and reproducibility to the users. 
Containers can provide the requirements of a program and execute it directly, without the overhead that comes with hypervisor-based approaches.
%Recently released ``Singularity-containers'' from \cite{kurtzer_2016}, it focuses the container approach to provide portable environments which leverage the migration of computational science to the cloud. 
``Singularity-containers'' from \cite{singularity} is a container-based approach which focuses on providing portable environments which could leverage the migration of computational science to the cloud. 
Singularity integrates seamlessly with existing workload managers such as Slurm, HTCondor or Torque; fact that could ease its adoption of HPC facilities. 
%can deploy an application or code with minor drawbacks on performance.

At the distributed systems and networks laboratory, at Universidad del Valle, we are working on the deployment  of container-based software infrastructures to support the research process on different areas of knowledge. 
We have tested diverse operating system-based virtualization technologies running single node and multi-node applications getting important results which show that this kind of virtualization is prime time ready to support research processes. 
This paper presents a set of benchmarks that stress different aspects such as compute, memory bandwidth, memory latency, network bandwidth, and I/O bandwidth. 
We will present and compare three container-based operating systems (Docker, LXC and Singularity) in section II. Then, we will describe the methodology used alongside the results in order to evaluate performance overhead of container-based technologies versus the bare metal in section III. Related works will be addressed in section IV. 
%The conclusion and future work are presented in Section V.
%
%% End of section
%

%% file: costechnologies.tex
\begin{figure*}[ht]
\centering
\includegraphics[width=0.8\textwidth]{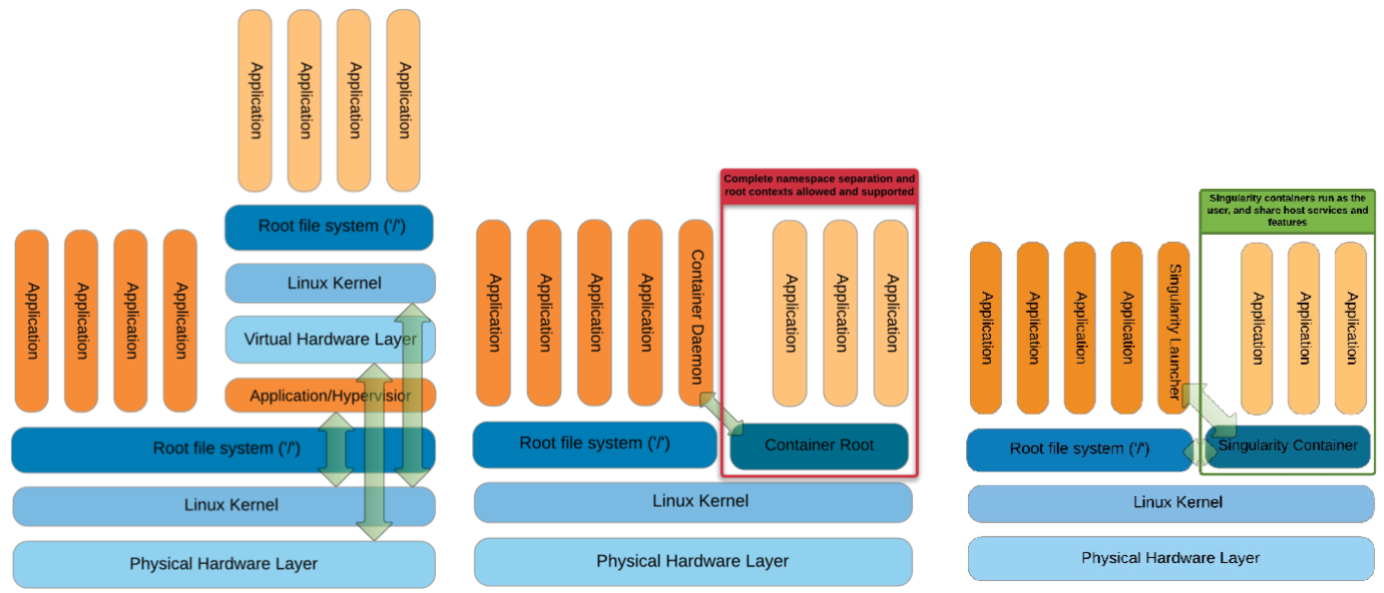}
\caption{(a) Architecture of hypervisor-based virtual environments. (b) Architecture of container-based virtual environments. (c) Singularity Architecture \label{fig:virtualization}}
\end{figure*}

\section{\label{sec:costech}Container-based Operating System Virtualization Technologies}
Containers are software components to enable the execution of applications on isolated environments.
Container-based operating systems (COS), also known as lightweight virtual machines, provide isolation of system resources (file system, network communications) in such a way that every container has its own set of processes ids, user identifiers, filesystem namespace and so on. 
Containers have a closer access to operating system services than their counterparts virtualization tools  such as native virtualization, paravirtualization and hypervisors.
Figure \ref{fig:virtualization}-a shows that containerized applications run almost at the same level of native applications. In contrast, classical virtualization approaches (Figure \ref{fig:virtualization}-b) propose several layers between applications in virtualized environments and the hardware where virtual machines are actually running. 
In fact, these layers impose a big overhead in virtualized applications when they are compared with  applications running on top of bare metal systems. 
Therefore COS technologies are now very attractive not only because they provide experimental reproducibility and platform portability but also because they exhibit a performance close to the performance exhibited on top of native environments \cite{ruiz:hal-01195549}.

%\begin{figure}[t]
%\begin{center}
%\includegraphics[width=0.4\textwidth]{figures/container-diagram}
%\caption{\label{fig:container} Container contexts run at same level of native applications.}
%\end{center}
%\end{figure}

%\begin{figure}
%\begin{center}
%\includegraphics[width=0.4\textwidth]{figures/fullvirtualization-diagram}
%\caption{\label{fig:fullvirtualization} Applcations in VMs are far from host hardware  }
%\end{center}
%\end{figure}

COS have being around for awhile and there are numerous implementations of it. On 2000, FreeBSD (4.0) featured the Jails system which focused on providing an isolated filesystem (an enhanced version of the $chroot$ command). Solaris goes a step further with its operating system OpenSolaris providing not only isolation services but also mechanisms related to snapshots and cloning. These aforementioned projects were mostly supported by BSD operating systems.
On 2005 OpenVZ was announced as a COS implementation for Linux systems. Despite it was an open source project there was not too much interest in the Linux community then it was barely included into the Kernel main stream. OpenVZ never gets enough track amongst Linux community.

LXC (Linux Containers) took advantage of the namespace concept. Different from previous approaches where file system isolation was provided, LXC extended the isolation property to users, processes and networking. On 2001, Linux supported the first file system namespace known as the mount namespace. Since then, other namespaces have been supported, UTS, IPC, PID, user and network namespaces. In addition to isolation, on 2006, Google project (process containers) implemented a functionality to limit the resource usage, e.g. CPU, memory, disk I/O, network). This project was later merged into the Linux kernel and it was named cgroups (control groups). From that, cgroups capabilities have been extended to firewalling and unified hierarchy, amongst others.

Docker released on 2013, was basically an additional layer on top of LXC exposing additional features such as mounted storage, network port redirection, and container catalog management. These features made Docker a prime time product for the industry.

Singularity is a project developed at Lawrence Berkeley National Laboratory (LBNL) and it is mainly focused on experimental reproducibility and isolation. 

%\href{https://www.cybera.ca/news-and-events/tech-radar/contain-your-enthusiasm-part-two-jails-zones-openvz-and-lxc/}{OpenVZ, Jails, LXC}

%Virtuozzo, VServer, Docker, Singularity.

%\href{http://www.slideshare.net/kolyshkin/openvz-virtuozzo-and-docker}{OpenVZ, Virtuozzo, Docker}

%\href{https://coreos.com/rkt/docs/latest/rkt-vs-other-projects.html}{rkt comparison}

%\href{https://stgraber.org/2016/03/11/lxd-2-0-introduction-to-lxd-112/}{LXC/LXD}

%\href{https://aucouranton.com/2014/06/13/linux-containers-parallels-lxc-openvz-docker-and-more/}{lxc,openvz,docker,parallels}
%
%% LXC
%
\subsection{LXC}
LXC is built on top of kernel namespaces which is a Linux kernel feature that isolates and virtualizes system resources such as processes, network, filesystems, network stack, among others\footnote{http://haifux.org/lectures/320/netLec8\_final.pdf}.  These capabilities allow a fully operational container-based environments exhibiting interesting features such as exposure of network services from containers, containers live migration, and a complete set of accountability mechanisms\cite{xavier2013performance}.  
In the networking context, LXC supports route- and bridge-based networking which allow the communication with the outside world but these features add a virtual network layer over the host which imposes an overhead to the network performance.
Mechanisms based on cgroups are used for restraining the amount of resources that a container can consume, e.g. CPU, memory, number of opened files and so on.
The container scheduling follows two level CPU scheduler which tries to promote fair scheduling among containers. 
First level scheduler determines which container will run, the second level determines which process in that container actually will run.
For I/O bandwidth there is also a two-level scheduling mechanism known as Completely Fair Queuing (CFQ) scheduler. 
Each container has a priority and inside of it an I/O bandwidth is given according to priorities.
%
%% Docker
%
\subsection{Docker}
Docker basically extends LXC with a kernel-and application level API \cite{bernstein2014containers} and mainly focusing on network service virtualization. 
Through the libcontainer library, Docker provides access to virtualization facilities provided by the Linux kernel along with some abstracted virtualized interfaces such as libvirt\footnote{https://libvirt.org/}, LXC and systemd-nspawn\footnote{https://www.freedesktop.org/software/systemd/man/systemd-nspawn.html}. 
The control over host's resources is provided through Control Groups (cgroups) thus it limits the amount of resources used by a container such as memory, disk space and I/O  \cite{wiki:docker}. 
Docker features a layered filesystem called AuFS (Advanced Multi-Layered Unification Filesystem) which allows to overlay one or more existing filesystems. When a process needs to create a copy, AuFS creates a copy of that file. This feature provides image versioning management and exposing base images to more specialized virtualized systems \cite{merkel2014docker}.
Docker has emerged as a key player in the virtualization field since it has being widely adopted in the industry and academy because it leverages infrastructure consolidation and exhibits a low resource footprint.
Docker has boosted the adoption of service oriented architectures (e.g. microservices \cite{fowler2014microservices}) because it ease the deployment of self-contained modules which are able to independently interact with third parties using well-known and widely adopted network protocols (e.g. web services). These service oriented architectures encourage the adoption of adaptable and extensible computational environments (e.g. workflows) accelerating the pace of scientific progress \cite{gil2007examining}.  
%
%% Singularity
%
\subsection{Singularity}
Singularity\cite{singularity} is another container-based approach developed at Lawrence Berkeley National Laboratory (LBNL)\cite{laytonContainer}. It was created with the idea of compute mobility in mind. 
Although Singularity uses namespaces, it is used basically for application portability instead of host virtualization. In other words Singularity virtualizes only what is necessary to achieve run-time application container and portable environments.
Singularity does not support user escalation or context changes therefore Singularity's container inherits permissions of the user who runs that container. 
Because it does not support context change then I/O operations flow directly between environments where those operations are happening reducing the operation overhead and execution times.
Singularity seamlessly integrates with diverse HPC environments and tools, e.g. resource managers, HPC file systems, GPUs, etc.
Singularity's design enables the utilization of vintage Container OS like RHEL 5 and also supports Docker-based images. 

\subsection{Comparing LXC, Docker and Singularity}

\begin{table}[htp]
\caption{Features of LXC, Docker and Singularity}\label{tab:comparison}
\def\arraystretch{1.5}
\begin{tabular}{|p{3cm}|c|c|c|}
\hline
\backslashbox{Feature}{COS} & \makebox[3em]{LXC} & \makebox[3em]{Docker} & \makebox[3em]{Singularity} \\
\hline
Support namespaces & Yes & Yes & Yes  \\
\hline
Support cgroups & Yes & Yes & No \\
\hline
Support port mapping & Yes & Yes & No \\
\hline
User  escalation & Yes & Yes & No \\
\hline
Unprivileged hardware access & No & No & Yes \\
\hline
API for applications and developers & Yes & Yes & No \\
\hline
Image Layering & No & Yes & Yes \\
\hline
Support snapshots & Yes & Yes & No (\footnotemark[1])  \\
\hline
Network interface                        & Host or Bridge & Bridge  & Host       \\
\hline
Default filesystem                        & Host (\footnotemark[2])       & AuFS    & ext3    \\
\hline
Access to host filesystem                  & Yes         & Yes & Yes    \\
\hline
Root daemon                               & Yes            & Yes     & No         \\
\hline
Registry/Repository for the images        & Yes            & Yes     & Yes        \\
\hline
Build a container from a file             & No             & Yes     & Yes        \\
\hline
HPC accommodations                         & No             & No  & Yes    \\
\hline
Keep modifications after restart          & Yes            & No      & Yes       \\
\hline
\end{tabular}
\end{table}
%
%% Comparison
%
Three COS technologies have been discussed.  Table \ref{tab:comparison} presents a summary of some features described above (versions compared are given in table  \ref{table:1}).  The authors would like to note that this table, and the following discussion, is relevant to the time of writing of this paper.
Docker has most of all those characteristics (e.g. user escalation, API for developers, versioning management), then it is a very handy tool for leveraging the development of sophisticated enterprise and research tools.
LXC is known as the predecessor of Docker. LXC has evolved in the meantime. One big evolution is LXD which is an API for LXC. For instance, it allows the live migration of containers between different LXD hosts. Therefore, users of LXC/LXD are not the same than the docker users (e.g. LXC is provided by the Proxmox virtualization server solution \footnote{https://www.proxmox.com}). Indeed, state full containers and full environment (like singularity) could be quite interesting features.
On the other hand, Singularity  exhibits a limited number of properties because it is mostly conceived for code mobility and high availability of resources.

In next section, HPC benchmarks were run against these COS technologies and preliminary results exhibit that the absence of some features positively affects the performance of these benchmarks.

\footnotetext[1]{However, a singularity image is only one file.}
\footnotetext[2]{Tight integration with ZFS.}

%namespaces (docker, lxd, singularity), cgroups (docker, lxd), map ports (docker, lxd), map directories (docker, lxd),

%% file: methodology.tex
\section{Methodology and Benchmarks}
This section studies the computational performance of COS technologies vs bare metal.
We performed several experiments with the current most popular COS implementations. Virtualization technologies and their versions are given in Table \ref{table:1}.

The performance experiments were executed at two facilities Universidad del Valle Cluster and Montpellier Bioinformatics Biodiversity cluster computing platform.

Configuration of computational nodes used on this work are as follows: CPU model Intel(R) Xeon(R) CPU E5-2683 v4 @ 2.10GHz(64-core node); Memory 164 GB DDR3-1,866 MHz, 72-bit wide bus at 14.9 GB/s on P244br and a HPE Dynamic Smart Array B140i Disk; OS Ubuntu 16.04 (64-bit) distribution was installed on the host machine.  

We used the industry reference HPL-Lapack benchmark to test CPU performance, and microbenchmarks to individually measure memory, network,  I/O and GPU overhead.

We know that results may vary significantly depending on the CPU architecture. versions of the kernel may introduce gains and losses of performance that would influence the results of experiments. Hence, we took care of compiling the same sha1sum binary for all benchmarks, using the host network for Singularity.

\begin{table}[h!]
\begin{center}
\def\arraystretch{1.5}
\begin{tabular}{ |p{3cm}|c| } 
 \hline
\centering Virtualization technologies & Versions  \\ 
 \hline
 Singularity 	& 2.2.1  \\ 
 \hline
 Docker 		& 17.03.0-ce, build 60ccb22  \\ 
 \hline
 LXC 			& 2.0.9  \\ 
 \hline
\end{tabular}
\end{center}
\caption{\label{table:1} Virtualization technologies and their versions}
\end{table}
%
%% Elapsed time for echo "Hello world"
%
\subsection{Time to execute a basic operation}
The basic operation includes the start-up of the container and the execution of the very basic and very well known command ``/bin/echo Hello World''. We used ``/usr/bin/time'' from the host to monitor it. From a native point of view, it always took 0.00 second. The three containers (LXC, Docker, Singularity) are minimalists and have been similarly built. The images are already present on the host. 

\begin{figure}[t]
\begin{center}
\includegraphics[width=0.5\textwidth]{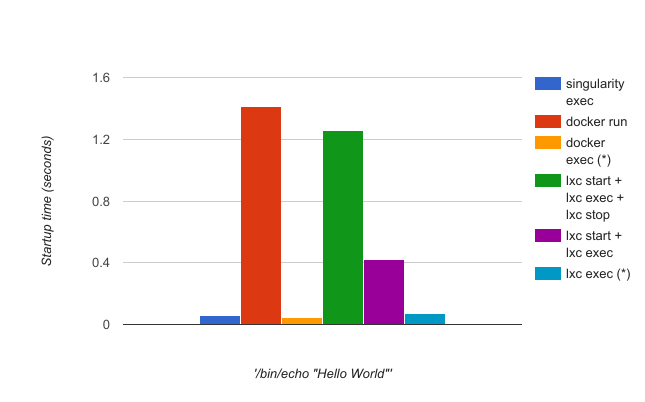}
\caption{\label{fig:8}Elapsed time (in seconds) for running "/bin/echo Hello Word".}
\end{center}
\end{figure}

We compared 6 operations fig \ref{fig:8} : the native ``/bin/echo Hello World'', and the same within ``singularity exec'', ``docker run'', ``docker exec'', ``lxc start + lxc exec'' and ``lxc exec''. The ``docker run'' command includes the boot, the execution and the shutdown of the container, while ``lxc exec'' and ``docker exec'' need a running container. So, we decided to add ``lxc start + lxc exec + lxc stop'' to the chart. Considering this graph, all the shutdown operations of a container are the slowest.

Then, we analyzed the strace\footnote{http://man7.org/linux/man-pages/man1/strace.1.html} outputs of the previous commands, and we did not notice any specific bottleneck. 
However, we observed a significant amount of ``futex'' (143) and ``rt\_sigprocmask'' (122) operations with ``docker run''.
These operations are usually dealing with synchronization mechanisms over threads when they are accessing shared resources e.g. shared memory regions. 
``docker run'' is an operation to create a new container (a.k.a. new running process) which requires to access and modify shared resources and data structures at kernel level.

%
%% HPL benchmark
%
\subsection{CPU performance}

In order to evaluate COS technologies for HPC we run the HPL-Benchmark \cite{petitet2004hpl} for a real vs virtual cluster as the ratio between the HPL benchmark performance of the cluster and the performance of a real environment formed with only one instance of same type, expressed as a percentage.

The benchmark were compiled using GNU C/C++ 5.4 and OpenMPI 2.0.2. We did not use any additional architecture- or instance-dependent optimizations. We used the SHA-1 hashes \cite{eastlake2001us} with the sha1sum program and checked the libraries with the ldd utility to ensure the binaries integrity. For the HPL benchmark, the performance results depend on two main factors: the Basic Linear Algebra Subprogram (BLAS) \cite{dongarra2002preface} library, and the problem size. We used in our experiments the GotoBLAS library, which is one of the best portable solutions, freely available to scientists. Searching for the problem size that can deliver peak performance is extensive; instead, we used the same problem size 10 times (10 N, 115840 Ns) for performance analysis. 

\begin{figure}[h]
\begin{center}
\includegraphics[width=0.5\textwidth]{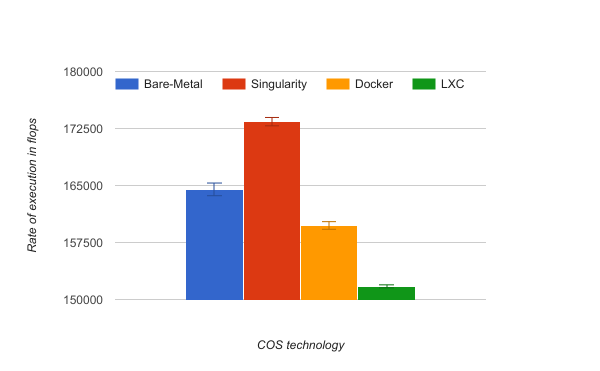}
\caption{\label{fig:1}Rate of execution for solving the linear system.}
\end{center}
\end{figure}

Figure \ref{fig:1} shows the performance of HPL-Benchmark. The Y axis is demonstrating the differences in technologies (that is why it doesn't goto zero). The LXC was not able to achieve native performance presenting an average overhead of 7.76$\%$, Docker overhead was 2.89$\%$, this could be probably caused by the default CPU use restrictions set on the daemon which by default each container is allowed to use a node’s CPU for a predefined amount of time. Singularity was able to achieve a better performance than native with 5.42$\%$ because is not emulating a full hardware level virtualization (only the mount namespace) paradigm and as the image itself is only a single metadata lookup this can yield in very high performance benefits.

%
%% IOzone benchmark
%
\subsection{Disk I/O performance}
The disk performance was evaluated with the IOzone benchmark \cite{iozone}. It generates and measures a variety of file operations and access patterns (such as Initial Write, Read, Re-Read and Rewrite). We ran the benchmark with a file size of 15GB and 64KB for the record size, under two(2) scenarios. The first scenario was a totally contained filesystem (without any bind or mount volume), and the second scenario was a NFS binding from the local cluster. 

\begin{figure}[t]
\begin{center}
\includegraphics[width=0.5\textwidth]{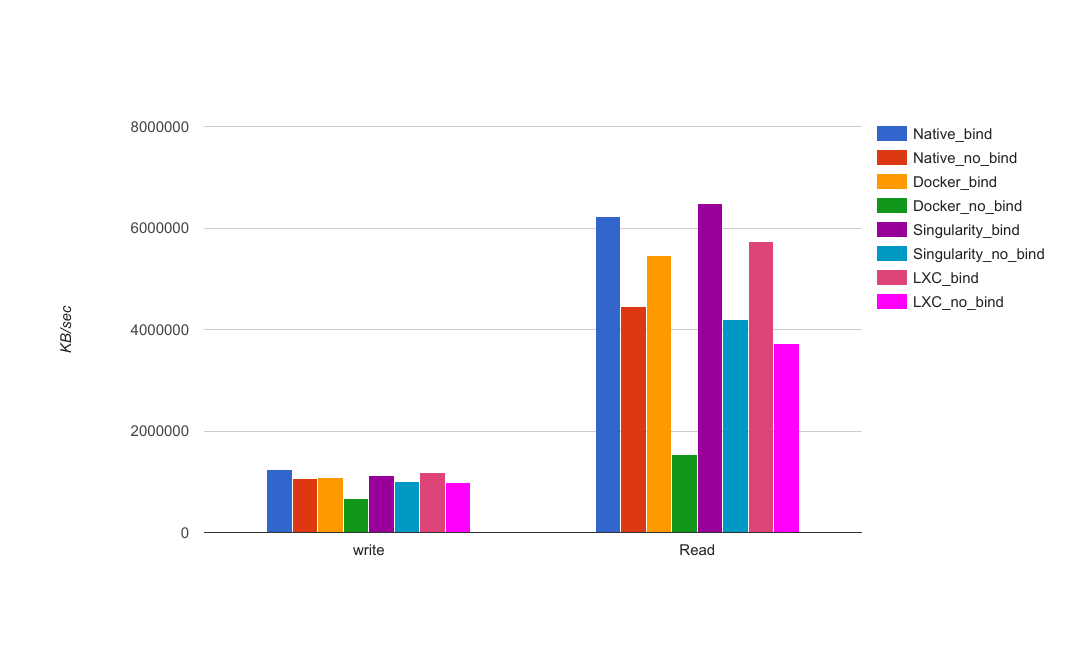}
\caption{\label{fig:2}IOzone benchmark write and read.}
\end{center}
\end{figure}

\begin{figure}[t]
\begin{center}
\includegraphics[width=0.5\textwidth]{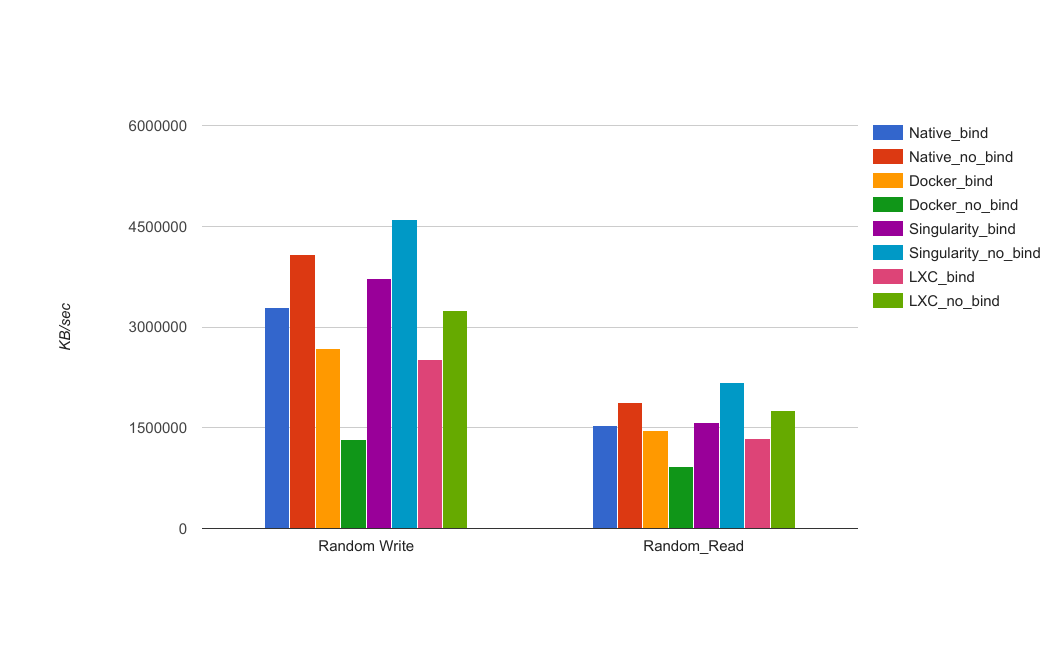}
\caption{\label{fig:3}IOzone benchmark random write and read.}
\end{center}
\end{figure}
A closer inspection in COS shown in Figure \ref{fig:2} reveals that both LXC and Singularity had similar results for write operations. For read operations, where the Singularity slightly reach the native performance, and LXC had an overhead of 16.39$\%$ against native. On the other hand, with Docker, we observed a lost of performance of 37.28$\%$ on write and 65.25$\%$ on read. Figure \ref{fig:3} shows the performance of random read and random write. We noticed a similar behavior than the read and write standard operations. Docker introduces a greater overhead on random I/O processes.
While LXC and Singularity filesystem implementations allows a better I/O performance, Docker advanced multi-layered unification filesystem (AUFS) has it drawbacks. When an application running in a container needs to write a single new value to a file on a AUFS, it must copy on write up the file from the underlying image. The AUFS storage driver searches each image layer for the file. The search order is from top to bottom. When it is found, the entire file is copied up to the container’s top writable layer. From there, it can be opened and modified.\cite{docker_aufs}
%
%% STREAM
%
\subsection{Memory performance}
The Memory performance on single node was evaluated with the STREAM application benchmark\cite{mccalpin1995sustainable}. It is a simple synthetic benchmark program that measures sustainable memory bandwidth (in MB/s) and the corresponding computation rate for simple vector kernels \cite{McCalpin2007}. The STREAM benchmark is specifically designed to work with datasets much larger than the available cache on any given system, so that the results are (presumably) more indicative of the performance of very large, vector style applications.  Performance evaluation is tight to the memory bandwidth of the system. Performance is therefore gated by memory bandwidth and not latency. The benchmark has four components: COPY, SCALE, ADD and TRIAD.

\begin{figure}[t]
\begin{center}
\includegraphics[width=0.5\textwidth]{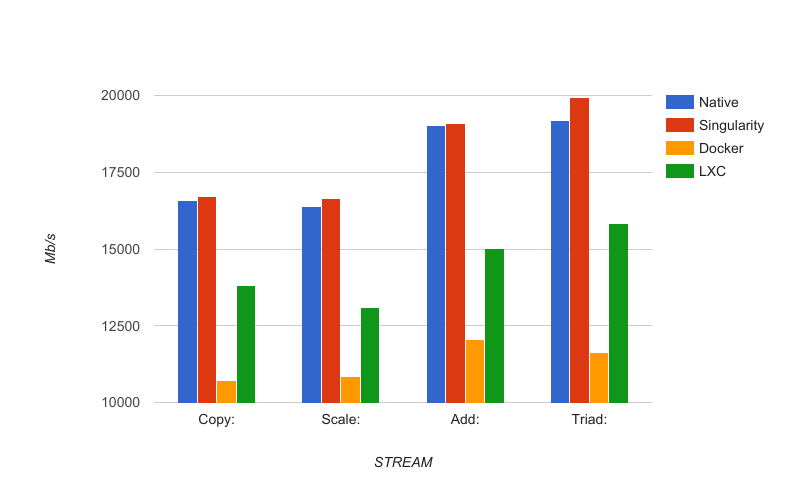}
\caption{\label{fig:4}STREAM benchmark results.}
\end{center}
\end{figure}

Results are presented in fig \ref{fig:4}. 

Figure \ref{fig:4} presents different performance for COS and native systems, for vector operation. This is due to the fact that container-based systems have no resource constraints and can use as much of a given resource as the host’s kernel scheduler will allow. The worst results were observed in Docker, which presented an average overhead of approximately 36$\%$ when compared to the native throughput.

%
%% OSU benchmarks (Network bandwidth and latency)
%
\subsection{Network bandwidth and latency performance}
For the MPI-level network evaluation we used the MVAPICH OSU Micro-Benchmarks 5.3.2 \cite{microosu} using a direct 10 Gbps Ethernet link between the nodes. We run point to point tests for measuring bandwidth and network latency.

Docker and LXC attaches all containers on the host to a bridge and connects the bridge to the network via NAT. We did not set any special network configuration for any technology, more than their native networking documented on each project web page (creating bridges for LXC and using Docker Swarm and creating an overlay network for Docker). Singularity needed no additional network configurations. 

\begin{figure}[t]
\begin{center}
\includegraphics[width=0.5\textwidth]{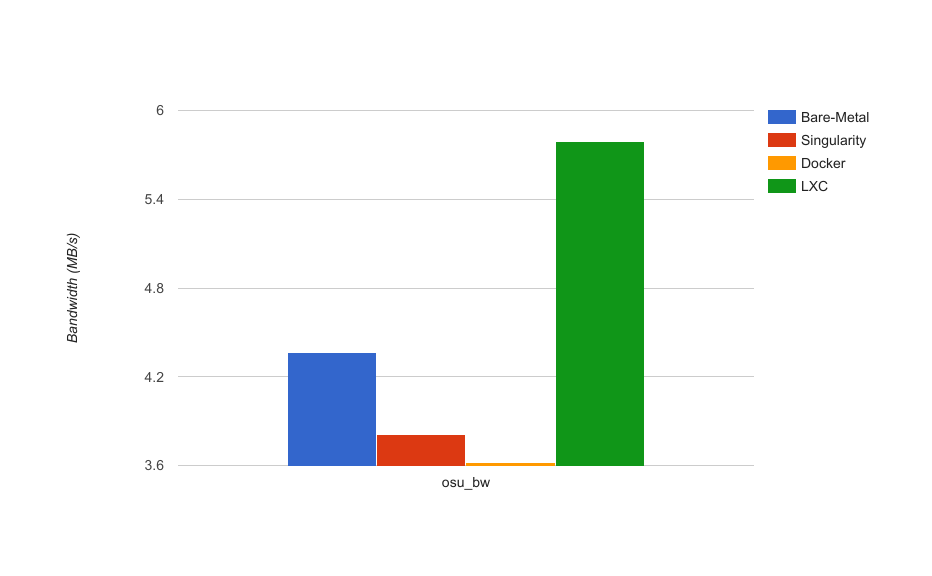}
\caption{\label{fig:5}OSU MPI bandwidth Test msgsize 4 MB.}
\end{center}
\end{figure}

\begin{figure}[t]
\begin{center}
\includegraphics[width=0.5\textwidth]{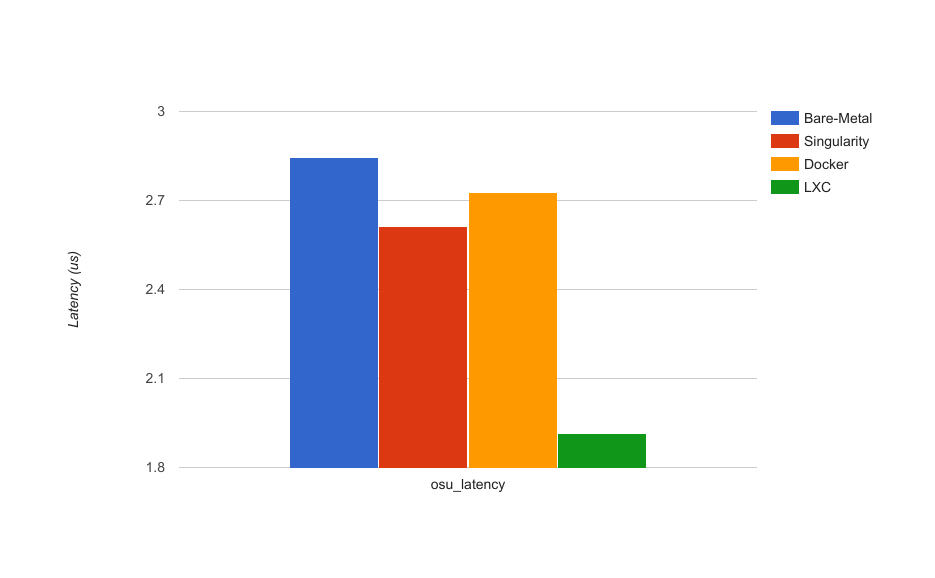}
\caption{\label{fig:6}OSU MPI Latency Test msgsize 1 byte.}
\end{center}
\end{figure}

Figure \ref{fig:5} shows the network bandwidth comparison for the COS. LXC has the best network scores with a great difference against the other two COS being evaluated. The singularity container showed a lower performance than the native implementation followed by Docker which presented the worst results. Its average bandwidth was 16.96$\%$ smaller than native. Figure \ref{fig:6} shows that LXC has less than 32$\%$ the network latency against native.  The worst bandwidth and latency was observed with Docker. These results can be explained due to different implementations of the network isolation of the virtualization systems. While Singularity container does not implement virtualized network devices, both Docker and LXC implement network namespace that provides an entire network subsystem. COS network performance degradation is caused by the extra complexity of transmit and receive packets (e.g. Daemon processes).
%
%% NAMD Benchmark
%
\subsection{GPU performance}
The performance studies were executed on a Dell PowerEdge R720, with 2*Intel(R) Xeon(R) CPU E5-2603 @ 1.80GHz (8 cores) and a NVIDIA Tesla K20M.\footnote{Kepler architecture\cite{lindholm2008nvidia}, GK110 Graphics processors, 2496 CORES, 208 TMUS, 40 Rops, 5120 MB Memory size, GDDR5 Memory type, 320 bit Bus width}. From a system point of view, we used Ubuntu 16.04.2 (64-bit), with NVIDIA cuda 8.0\cite{kirk2007nvidia} and the NVIDIA driver version 375.26. The virtualization technologies and their versions are given in Table \ref{table:2}.
\begin{table}[ht!]
\begin{center}
\def\arraystretch{1.5}
\begin{tabular}{ |p{3cm}|c| } 
 \hline
\centering Virtualization technologies & Versions  \\ 
 \hline
 Singularity 	& 2.2.1  \\ 
 \hline
 Docker 		& 17.03.0-ce, build 60ccb22  \\ 
 \hline
 LXC 			& 2.0.9  \\ 
 \hline
\end{tabular}
\end{center}
\caption{\label{table:2} Virtualization technologies and their versions}
\end{table}

In order to evaluate COS technologies for GPU-HPC, we used the NAMD (NAnoscale Molecular Dynamics) \cite{kale2011namd} program, as a benchmark tool. We ran those GPU benchmarks on a Tesla K20m with ``NAMD x86\_64 multicore CUDA version 2017-03-16'' [on the stmv dataset ( 1066628 Atoms )], using the 8 cores and the GPU card, without any specific additional configuration, except the use of the ``gpu4singularity''\footnote{https://github.com/NIH-HPC/gpu4singularity} code for Singularity and the ``nvidia-docker''\footnote{https://github.com/NVIDIA/nvidia-docker} tool for Docker. For a real vs virtual cluster, the ratio is printed in the log as "days/ns" (lower is better).

\begin{figure}[t]
\begin{center}
\includegraphics[width=0.5\textwidth]{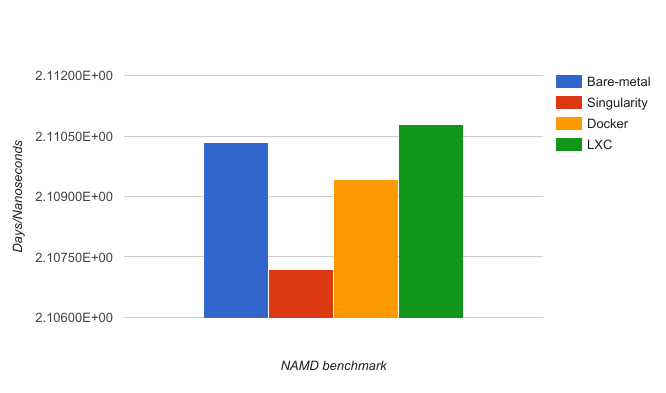}
\caption{\label{fig:7}Tesla K20m benchmarks on NAMD.}
\end{center}
\end{figure}
Figure \ref{fig:7} shows the performance of NAMD-Benchmark. The Y axis is in ``days/ns'' (the lower the better). LXC was not able to achieve native performance. Docker  achieved a better performance than native, which can be explained on the work that Nvidia is doing to build cloud-native gpu applications. Nevertheless, Docker does not natively support NVIDIA GPUs with containers \cite{nvidia_docker}. Singularity was able to achieve a better performance than native given that it provides native gpu support \cite{singularity}.
%
%% Github  reference
%
\subsection{Source Code}
The scripts to run the experiments from this paper are available
at \href{https://github.com/ArangoGutierrez/containers-benchs}{https://github.com/ArangoGutierrez/containers-benchs}

%% file: related_work.tex
\section{Related work}
Some papers have explored the overhead of container-based virtualization tools as presented by \cite{felter2015updated,xavier2013performance,kozhirbayev2017performance}. Mostly they compare the performance overhead of COS versus classic Virtual machine technologies (e.g KVM, LinuxV-server). They all agree that the current resource management implementation for LXC and Docker, lead to poor isolation and security. 

Containers are proving to be an extremely valuable technology for scientific research, delivering benefits such as portability and reproducibility to scientific users. COS can emulate a single program and can be executed directly, with less overhead that with running a virtual machines. Indeed, some works already described COS technologies in a scientific use case \cite{moreews2015bioshadock,belmann2015bioboxes,o2017dockstore,di2017nextflow}.
Despite the advantages offered by container technologies, the implications for scientific computing, including HPC, are still unclear, although there are already some initiatives like Singularity, Shifter\cite{jacobsen2015contain} (See also benchmarks on Cray systems for Shifter \cite{bahls2016evaluating}), Charliecloud\cite{priedhorsky2016charliecloud}, cHPC\cite{weidner2017rethinking} or Docker Universe Applications in HTCondor\footnote{http://research.cs.wisc.edu/htcondor/manual/v8.4/2\_12Docker\_Universe.html}.

Many core technologies (a.k.a. GPU cards) are widely deployed on private data centers and more recently they have been exposed as services in the cloud computing landscape in order to attend the increasing growth for computational power in different fields of research, e.g. bioinformatics \cite{lee2013gpu}, storage processing \cite{Sun:2012:GHG:2367589.2367595} and multimedia \cite{zhu2011multimedia}, among others. 
Virtualization technologies and their impact on GPU technologies has not been broadly studied because the challenges exhibited to virtualize the functionality of GPUs and to support GPU passthrough.
J.P. Walters et al.\cite{6973796} run non-standardized benchmarks for KVM, Xen, VMWare and LXC.
Using GPU Passthrough, these virtualization technologies are put under test running CUDA and OpenCL applications.
Preliminary results exhibit a penalty over 10\% on Xen and KVM. 
VMWare exhibited an irregular performance and LXC showed a closest performance to the native case.

To the best of our knowledge, there are no similar publications to this one. 
In particular, our work assesses three COS technologies using standardized benchmarks.
This approach gives a preliminary approach to characterize the assessment of COS and virtualization technologies in general.

There is a little research effort around container-based solutions for heavy HPC applications. Performance evaluations on literature usually does not put under test  the HPL-LAPACK benchmark, instead a version of HPL-LINPACK, where matrix size could impact on the final result (e.g CPU cycles). 
Reports like \cite{felter2015updated} used a compiled version of LINPACK from INTEL, here we compiled the binary inside of each COS, to replicate a normal work flow, when running on a HPC cluster. 
Moreover, HPL results may vary significantly depending on the CPU architecture.

%% file: conclusions.tex
%
%% Where results become ideas!
%
\section{Conclusions}
This paper presented a performance comparison of container-based virtualization tools (Docker, LXC, Singularity) against bare-metal. According to our results, we observed that Singularity containers are usually more suitable for HPC implementation than Docker or LXC. 
From a network point of view, LXC is very efficient, however, not all namespaces are equal, and Singularity does swap out the user namespaces. Therefore, if the container have more efficient libraries than the host, the Singularity solution could yield a performance increase, while LXC and Docker control their resources by cgroups namespace, which results on a overhead for CPU intensive processes. Besides, Singularity optimizes HPC-specific libraries like CUDA or OpenMPI.
For GPU applications, we recommend the implementation of Docker and Singularity to deploy on HPC clusters, or in the cloud. CUDA accelerated machine learning projects have already started offering Dockerfiles in order to run those applications over a container ready system. \cite{nvidia_docker}

For I/O-intensive workloads, Container images can be much more optimal then running against shared storage (even when the container image exists on that remote storage). That is normal, as this is the same principle as cache or even the HPC scratch, that is to say a way to have data close to the process. Moreover, we would avoid the use of the standard Docker-based solutions (AUFS), due to overhead issues.

Concerning small tasks like ``Hello World'', or even a more consistent memory job (see STREAM results), one more time, we would avoid the use of Docker, except if the container is already running on the host. That could lead to a real problem in a HPC system where the image needs to be downloaded everywhere and then, started, before being executed. Contrary to Singularity, where you have only one file, which can be shared through the network or can be stored on a distributed filesystem. Furthermore, a distributed filesystem can be significantly impacted by the metadatas accesses, thus Singularity can limit that problem.

Singularity blocks privilege escalation within the container to avoid users of having root access. Docker instead must be isolated thus will preclude access to high performance networks (e.g. InfiniBand) and optimized storage platforms.

Considering our container-based results, COS and particularly Singularity are a good alternative to overcome the virtualization overhead issues. COS environments can be used on mixed environments where HPC and HPSS are required. In order to take the HPC scenario, COS technologies must focus on: container overhead, container technology and architecture concerns (e.g. privilege escalation and network/file system access), and workflow compatibility.

%% file: acknowledgments.tex
%
%% Always say thanks 
%
\section{Acknowledgments}
This work tests where heavily run on the Engineering faculty Cluster, University of Valle.
This work also largely benefited from the Montpellier Bioinformatics Biodiversity cluster computing platform.
We would also like to thank David Godlove from the NIH Biowulf Cluster for its code on GPU.